\documentclass[%
 reprint,
 article
 amsmath,amssymb,
 aps,
]{revtex4-1}
\usepackage{amsmath}
\usepackage{dcolumn}
\usepackage{graphicx}
\usepackage{float}

\begin{document}

\preprint{APS/123-QED}

\title{Highly anisotropic superconducting gaps and possible evidence of antiferromagnetic order in FeSe single crystals}

\author{Guan-Yu Chen, Xiyu Zhu, Huan Yang, and Hai-Hu Wen}\email{hhwen@nju.edu.cn}

\affiliation{National Laboratory of Solid State Microstructures and Department of Physics, Collaborative Innovation Center of Advanced Microstructures, Nanjing University, Nanjing 210093, China}

\begin{abstract}
Specific heat has been measured in FeSe single crystals down to 0.414 K under magnetic fields up to 16 T. A sharp specific heat anomaly at about 8.2 K is observed and is related to the superconducting transition. Another jump of specific heat is observed at about 1.08 K which may either reflect an antiferromagnetic transition of the system or a superconducting transition arising from Al impurity. We would argue that this anomaly in low temperature region may be the long sought antiferromagnetic transition in FeSe. Global fitting in wide temperature region shows that the models with a single contribution with isotropic s-wave, anisotropic s-wave, and d-wave gap all do not work well, nor the two isotropic s-wave gaps. We then fit the data by a model with two components in which one has the gap function of $\Delta_0(1+\alpha cos2\theta)$. To have a good global fitting and the entropy conservation for the low temperature transition, we reach a conclusion that the gap minimum should be smaller than 0.15 meV ($\alpha$ = 0.9 to 1), indicating that the superconducting gap(s) are highly anisotropic. Our results are very consistent with the gap structure derived recently from the scanning tunneling spectroscopy measurements and yield specific heat contributions of about 32\% weight from the hole pocket and 68\% from the electron pockets.

\begin{description}
\item[Subject Areas]Condensed Matter Physics, Superconductivity
\end{description}
\end{abstract}

\maketitle
The iron-selenium (FeSe) is one of the iron based superconductors with the simplest structure \cite{WuMKPNAS2008}. The later effort in enhancing the superconducting transition temperature from about 8.5 K to 37 K by pressure was encouraging \cite{Imai37K}. However, the interest has been revived recently by exploring the phase diagram under pressure \cite{Phase1Kothapalli,Phase2,Phase3}. In the normal state under ambient pressure, there is a structural transition from tetragonal to orthorhombic at about 90 K \cite{BoehmerPRB2013}. This transition has been proved to be accompanied by the formation of the nematic electronic state \cite{TanatarPRL2016}. In contrast, however, there is no evidence of antiferromagnetic long range order found in the system although the normal state is dominated by very strong spin fluctuations \cite{ZhaoJunNatMat2015}. This may help to resolve the disputes about the origin of the nematicity \cite{BoehmerPRL2015}. Under pressure, together with the enhancement of superconducting transition temperature, an antiferromagnetic (AF) order appears \cite{Phase1Kothapalli,Phase2,Phase3}. However, it remains unknown how the AF order extends to the superconducting state in the low pressure region. It might be possible that this AF order is hidden under the superconducting dome and most measurements have been undertaken above the possible AF transition temperature. Angle resolved photoemission (ARPES) \cite{ShenZX,DingH} and scanning tunneling spectroscopy (STS) \cite{Hanaguri} have revealed the existence of both electron and hole pockets with very small area, showing the approximate semi-metal behavior. The closeness of the band edge to the Fermi energy, or the comparable scales of the Fermi energy $E_F$ and the superconducting gap $\Delta$, suggests a possible BCS-BEC crossover \cite{MatrusdaPNAS}. The pairing order parameter has been detected by many experimental techniques. The first STS experiment on the sister system FeSe$_{1-x}$Te$_x$ has revealed the sign reversal s$^\pm$ gap. However, in FeSe thick films, a V-shaped spectrum has been observed suggesting the d-wave gap structure \cite{XueQK2010Science}. Thermal conductivity measurement has been carried out which suggests a nodeless gap but with a very small gap minimum \cite{TaillefferPRL2016}. In this paper, we report low temperature specific heat on the FeSe single crystal. We have fitted the experimental data with variable kinds of gap structures. Our results suggest the superconducting order parameter with two components of highly anisotropic gaps. In addition, a clear step like specific heat anomaly at about 1.08 K gives the possible evidence of an AF order in low temperature region.

\begin{figure}
\includegraphics[width=9.4cm]{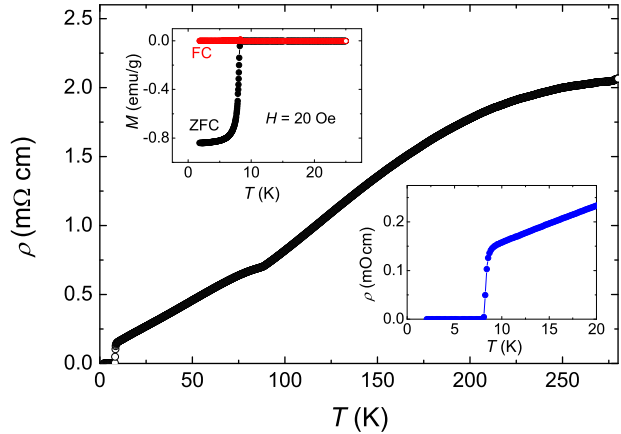}
\caption{\label{fig1} Main panel: Temperature dependence of resistivity of one FeSe single crystal sample at zero magnetic field. The upper-left inset shows the magnetization measured in the ZFC and FC modes with external magnetic field of 20 Oe. The right-bottom inset shows the enlarged view of the resistive transition. A superconducting transition at about 8.2 K is obvious.}
\end{figure}

\section{EXPERIMENT}
The FeSe single crystals used for this study were grown by the chemical vapor transport method using Fe and Se powder as the starting materials \cite{BoehmerPRB2013}. The mixture of Fe$_{1.04}$Se, KCl and AlCl$_3$ with the ratio of 1:2:4 were put at one end of the quartz tube in a glove box filled with argon. We then sealed the quartz tube and place it into a horizontal tube furnace with tunable temperature gradient. The furnace was heated up to 430 $^\circ$C and kept for 30 hours, then by adjusting the program the temperature at the end without reactant was tuned to about 370 $^\circ$C, in order to establish a temperature gradient. The tube was sintered with this temperature gradient for 6 weeks, and finally FeSe single crystals were grown at the cold end of the tube. The magnetization was measured by using a superconducting quantum interference device (SQUID-VSM 7T, Quantum Design). The resistivity was measured with a physical property measurement system (PPMS 16T, Quantum Design) by the standard four probe method. The specific heat was measured with thermal-relaxation method by an option of the PPMS with a He3 insert. This facility allows us to measure specific heat down to 0.414 K. During the specific heat measurement, magnetic fields up to 16 T was applied parallel to the c-axis of the crystal.

\begin{figure}
\includegraphics[width=8.5cm]{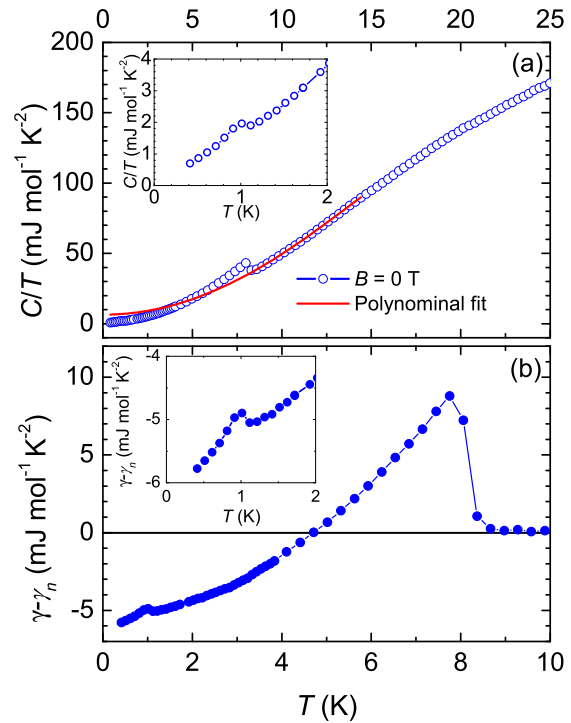}
\caption{\label{fig2} (a) The circles represent the raw data of specific heat coefficient $C/T$ for the FeSe single crystal. A major specific heat jump is observed at around 8.2 K. The red solid line is a fit to the data of C/T vs. T above T$_c$ based on the Debye model (Eq.1). (b) The specific heat coefficient (C-C$_n$)/T vs. T. The insets in (a) and (b) show the enlarged views of the same data in each panel.}
\end{figure}

\section{RESULTS AND DISCUSSION}

\begin{figure}
\includegraphics[width=8cm]{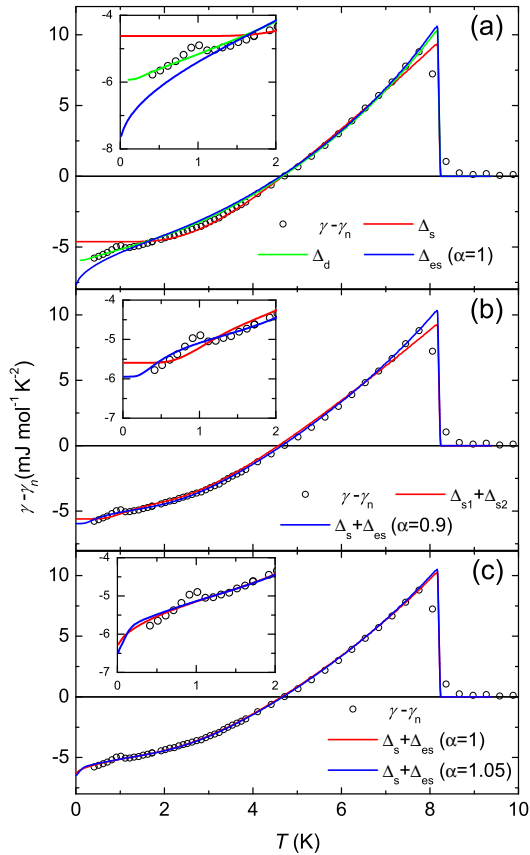}
\caption{\label{fig3} Raw data of the electronic specific heat coefficient vs. temperature (symbols), and the fitting curves with different combinations of gaps. The expression for $\Delta_{es}$ is $\Delta_0(T)(1 + \alpha cos2\theta)$. (a) The red, green and blue lines show the global fitting to the data with a single isotropic $s$-wave, single $d$-wave, single extended $s$-wave, respectively. (b) The red line shows the fitting with two isotropic $s$-wave gaps. The blue line shows the fitting with two components, one isotropic s-wave and one extended s-wave ($\alpha$ = 0.9). (c) The red line shows the fitting with two components: one isotropic s-wave and one extended s-wave ($\alpha$ = 1). The blue line shows the fitting with two components: one isotropic s-wave and one extended s-wave ($\alpha$ = 1.05). The insets in (a), (b) and (c) show the corresponding enlarged views of the fitting result below 2 K, so the scales and labels are the same as those in each main panel.  The second transition here serves as a very nice indicator for checking the appropriate gap structure.}
\end{figure}

\subsection{Basic characterization of the sample}
The temperature dependence of resistivity $\rho(T)$ at zero magnetic field is shown in Fig.~\ref{fig1}. We can see a clear kink at about 88 K, which is related to the structural transition from tetragonal to orthorhombic phase and the nematic transition as well. The onset of the superconducting transition $T_c^{onset}\approx$ 8.7 K (defined by the crossing point of the extrapolated lines of the normal state and the steep transition part), and the system realizes zero resistivity state at $T_{c0}\approx$ 8.1 K, which can be seen in the right-bottom inset of Fig.~\ref{fig1}. The transition width $\Delta T_c$, which is defined as  $\Delta T_c$ = $T_c^{onset}-T_{c0}$, is 0.6 K. And the residual resistivity ratio (RRR), which is determined by the ratio of $\rho(300K)/\rho(T=0K)$, is about 25.2, where $\rho(T=0K)$ is obtained by linearly extrapolating the normal state resistivity down to zero temperature. The temperature dependence of zero-field-cooled (ZFC) and field-cooled (FC) magnetization at 20 Oe is shown in the upper-left inset of Fig.~\ref{fig1}. The superconducting volume calculated from the magnetization data is larger than 100\% (due to the demagnetization effect), indicates the bulk superconductivity of our sample. The large RRR value and superconducting volume both confirm the high quality of our samples.

\subsection{Measurement on specific heat and fitting with different gap structures}
Specific heat is sensitive to the quasiparticle density of states (DOS) at the Fermi energy, so it is a useful way to detect superconducting gap structure at low temperatures. In iron based superconductors, specific heat measurements have been done in plenty systems \cite{HardyBaK122,MuGangBaK122,HardyK122} showing the multigap feature. In order to study the superconducting gap structures of the FeSe system, we have measured the specific heat of an FeSe single crystal and the temperature dependence of specific heat at zero field is presented in Fig.~\ref{fig2}(a). A sharp jump of specific heat coefficient can be seen at about 8.2 K, which is corresponding well to the superconducting transition detected by resistivity and magnetization measurements. Another anomaly occurs at about 1.08 K. The second transition at low temperature may reflect the possible antiferromagnetic transition, which have not been observed among previous researches. However, since AlCl$_3$ have been used during the progress of crystal growth, the possibility that the second jump is arisen from Al impurity ($T_c$ = 1.17 K) could not be excluded. But we will argue that the latter is unlikely. The residual specific heat $\gamma_0=C/T|_{T\rightarrow 0}$ determined by extrapolating the data of $C/T$ down to 0 K is negligible, which seems to against the possibility of any nodal gaps in the system, and is consistent with the recent thermal conductivity measurements \cite{TaillefferPRL2016}.

It is known that the specific heat consists both of the electronic and phonon contributions. Since the phonon contribution will prevail over the electronic part in moderate temperature region, it is very important to extract the superconducting electronic specific heat. Thus the data of the normal state above $T_c$ is fitted by the equation of Debye model which reads as

\begin{equation}
C_n/T =\gamma_n+\beta T^2 +\eta T^4,\label{eq1}
\end{equation}

where $\gamma_n$ is the normal state electronic specific coefficient, or called as the Sommerfeld coefficient and $\beta T^2 +\eta T^4$ are the phonon contributions according to the Debye model. The fitting curve is shown by the red solid line in Fig.~\ref{fig2}(a). The fitting function yields $\gamma_n$ = 6.4 mJ/mol$\cdot K^2$, $\beta= 0.44 mJ/mol\cdot K^4$ and $\eta= -0.00005 mJ/mol\cdot K^6$. The Debye temperature estimated here is about 206 K, which can be obtained by using the equation $\Theta_D = (12\pi^4 k_B N_A Z/5\beta)^{1/3}$, where $k_B$ is the Boltzmann constant, $N_A$ is the Avogadro constant and Z is the number of atoms in one unit cell. Our $\gamma_n$ and $\Theta_D$ are close to the values of earlier studies \cite{J.Y.LinPRB2011}. The specific heat jump at $T_c$, namely $\Delta C/\gamma_n T_c$ with $\Delta C$ estimated through entropy conservation near $T_c$ is about 1.6. This is larger than 1.43 predicted by the Bardeen-Cooper-Schriefer (BCS) theory in the weak coupling limit, which may indicate moderate strong coupling in FeSe. The temperature dependence of superconducting electronic specific heat is plotted in Fig.~\ref{fig2}(b), which is obtained by subtracting the normal state data $C_n/T$. There is a very small tail above $T_c$, which may indicate the narrow  superconducting fluctuation region in FeSe (less than 1 K). The small jump at around 1.08 K can be clearly seen in the insets of Fig.~\ref{fig2} (a) and (b). A similar anomaly was also seen in earlier report \cite{Lin.Jiao2016arXiv}, which however exhibits like a shoulder around 2 K there, and the authors attributed this anomaly to a very small second superconducting gap. Since the low temperature anomaly observed here is a clear jumping step, not a knee like, this excludes the possibility that it is a second superconducting gap of the FeSe system, since a multi-band fitting using the same $T_c$ would give rise to a "knee" or "hump" here, not as a sharp step. We would like to attribute this step to either the antiferromagnetic order transition, or the impurity of Al which has a $T_c$ of about 1.17 K. We will address to this issue later.

\begin{table*}[htb]
\caption{\label{tab:table1}
The parameters derived from fitting to the data with different models}
\begin{ruledtabular}
\begin{tabular}{cccccccc}

 &$\Delta_1(meV)$ & $Fraction-1$ &$\Delta_2(meV)$ & $Fraction-2$ &  $\alpha$ \\
\hline
isotropic $s$-wave & 1.5 & 100\%  & - & - &  - \\
$d$-wave & 1.95 & 100\%   & - & - & - \\
extended $s$-wave    & 1 & 100\% & - & - &  1 \\
two isotropic $s$-wave    & 1.5 & 80\% & 0.45 & 20\% & - \\
isotropic $s$-wave and extended $s$-wave    & 1.45 & 47\% & 1.13 & 53\% &  0.9 \\
isotropic $s$-wave and extended $s$-wave    & 1.5 & 37.5\% & 1 & 62.5\% &  1 \\
isotropic $s$-wave and extended $s$-wave    & 1.5 & 36\% & 1 & 64\% & 1.05 \\
\end{tabular}
\end{ruledtabular}
\end{table*}

The structure of superconducting gap is a significant issue in determining the pairing mechanism, while, as far as we know, there is no consensus yet concerning whether nodes exist in the superconducting gap of FeSe or not. The STS measurements \cite{XueQK2010Science,Hanaguri} suggest that there might be nodes in this system, while the thermal conductivity measurement \cite{TaillefferPRL2016} and the specific heat measurement \cite{Lin.Jiao2016arXiv} support a nodeless gap. In order to obtain more information about the gap structure of FeSe, we use BCS formula to fit the electronic specific heat in superconducting state, the formula based on BCS theory is shown below,

\begin{eqnarray}
\gamma_\mathrm{e}=\frac{4N(E_F)}{k_BT^{3}}\int_{0}^{+\infty}\int_0^{2\pi}\frac{e^{\zeta/k_BT}}{(1+e^{\zeta/k_BT})^{2}}\nonumber\\
\nonumber\\
(\varepsilon^{2}+\Delta^{2}(\theta,T)-\frac{T}{2}\frac{d\Delta^{2}(\theta,T)}{dT})\,d\theta\,d\varepsilon,
\end{eqnarray}

where $\zeta=\sqrt{\varepsilon^2+\Delta^2(T,\theta)}$, the angle dependence of the gap is entered through $\Delta (\theta)$ here and the temperature dependence is generated from the BCS gap equation, $\varepsilon=\hbar^2k^2/2m$ is the kinetic energy of electrons counting from the Fermi energy. For the fitting to one gap, we just use above formula. While for the fitting with two gaps, we use a linear combination of two contributions, each one has a gap. Variable different gap structures have been used to fit our data: single $s$-wave gap $\Delta(T, \theta) = \Delta_0(T)$, single $d$-wave gap $\Delta(T, \theta)= \Delta_0(T)cos2\theta$, single extended $s$-wave gap $\Delta(T, \theta)= \Delta_0(T)(1 + \alpha cos2\theta)$, mixture of two isotropic $s$-wave gaps  $\Delta(T, \theta) = \Delta_1(T)+\Delta_2(T)$ and mixture of an $s$-wave gap and an extended $s$-wave gap. The $\alpha$ in the expression of extended $s$-wave is the parameter that represents anisotropy. One can see that $\alpha$ = 0 corresponds to the case of an isotropic $s$-wave gap, while $\alpha$ = 1 corresponds to a zero gap minimum. The optimized fitting parameters of different models are listed in Talbe~\ref{tab:table1} and the fitting curves are shown in Fig.~\ref{fig3}(a)-(c). The insets in Fig.~\ref{fig3}(a)-(c) provide the enlarged views below 2 K, so we can check the fitting results at low temperatures. Although the origin of the second jump at 1.08 K remains unknown, it will serve as a very nice indicator for checking the appropriate gap structure through the entropy conservation of this transition, especially about the gap minimum. In Fig.~\ref{fig3}(a), we show the fitting by using three cases of single gap, namely a single isotropic $s$-wave, single $d$-wave and singe extended $s$-wave. The formula used for the extended $s$-wave naturally has a two-fold symmetry if $\alpha$ is not zero, this form is chosen because it is well-known that the FeSe system has a nematic state. One can see that both of the single isotropic $s$-wave model and the single extended $s$-wave model ($\alpha$ = 1) are not possible to describe the experimental data, judging both from the global fitting quality in wide temperature region and the low temperature part. However, it seems a single $d$-wave fits the data much better. This may suggest that the gap should be highly anisotropic in order to get a global fitting in wide temperature region. A closer scrutiny finds that the $d$-wave fitting still has some deviation to our data between 2 K and 4 K. Combining with the fact that the residual specific heat coefficient in the zero temperature limit is negligible, we rule out the possibility of $d$-wave gap(s).  These results indicate that a model with only one component/gap is not enough to describe the data. This is reasonable since in the realistic case, there are at least two contributions, one from the hole pocket and one from the electron pocket. The model with two isotropic $s$-wave gaps is also tried and shown by the red solid line in Fig~\ref{fig3}(b), which shows a very poor fitting near $T_c$ and the deviation in low temperature is also clear. Therefore isotropic $s$-wave gap(s) with either single or double contributions cannot fit the data, and anisotropic gaps should be taken into account. Fig.~\ref{fig3}(b) and (c) show also fitting results by using the gap model with the mixture of an $s$-wave gap and an extended $s$-wave gap. Three different values of $\alpha$ have been tried to fit our data. All of these models work well above 2 K when $\alpha$ ranges from 0.9 to 1.05. However, these curves behave quite differently at low temperatures below 2 K. For the case of $\alpha$ = 0.9, as shown by the inset of Fig.~\ref{fig3}(b), the fitting is relatively better and the low temperature part seems also okay in terms of entropy conservation about the low temperature anomaly. When $\alpha$ = 1, the bending down of the fitting curve at low temperatures seems a little too much to satisfy the entropy conservation. For the case of $\alpha$ = 1.05, the bending down is simply too much without possibility of the conservation of the entropy of the low temperature anomaly. For the case of $\alpha$ = 0.8 or below, we see a flattening of the data starting at higher temperatures, which makes no case for the entropy conservation for the low temperature anomaly. Therefore, in order to have a good global fitting in wide temperature region and the conservation of entropy around the low temperature anomaly, we reach the conclusion that $\alpha$ locates around 0.9. Thus we can conclude that, the gap in FeSe should be highly anisotropic. Taking $\alpha$ = 0.9 to 1, we believe that the minimum of the gap is about 1.5$\times(1-0.9)$ = 0.15 meV or smaller. We must mention that, what we used here for the fitting are two components with one isotropic $s$-wave and one extended $s$-wave. One can also use a model with two extended $s$-wave gaps, but that requires more fitting parameters, which would give more uncertainty. About judging to what extent the gap anisotropy is, fitting to specific heat by using one $s$-wave plus an extended $s$-wave, or two extended $s$-wave gaps gives no big difference.

\begin{figure}
\includegraphics[width=8.5cm]{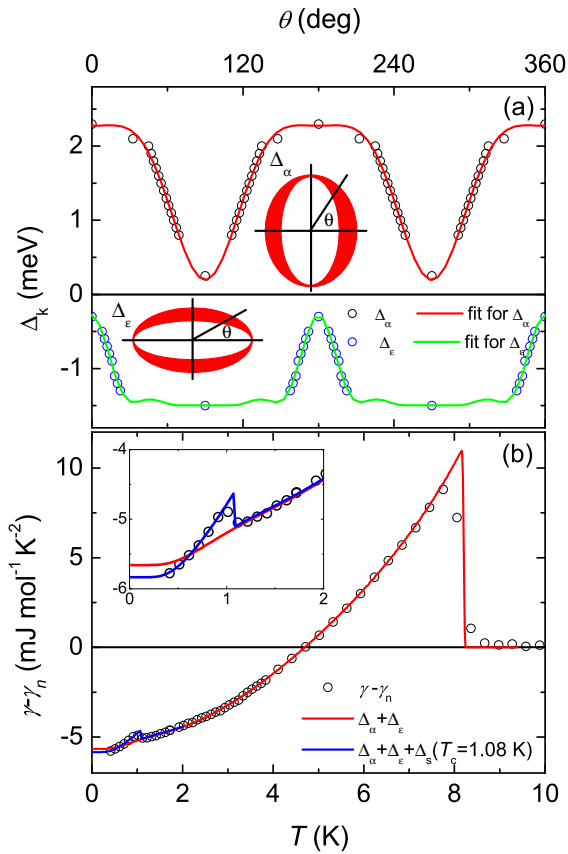}
\caption{\label{fig4} Fitting the specific heat data by using the gaps derived from STS measurements. (a) Angle dependence of the superconducting gap (symbols) and the fitting curves for the gaps at hole pocket or $\alpha$-band(red) and electron pocket or $\epsilon$-band(green). The insets illustrate the gap structure (the width of the filled area) on top of elliptic like Fermi surfaces and the definition of $\theta$. (b) The symbols represent the electronic specific heat $\gamma-\gamma_n =[C(T)-C_n(T)]/T$. The red line shows the fitting to the data with the gap structure displayed in (a). The blue line shows the fitting to the data with the gap structure in (a) plus a small $s$-wave gap with $T_c$= 1.08 K. The inset shows an enlarged view of the same data with the same scales and labels.}
\end{figure}

\subsection{Fitting with the gaps determined by STS experiment}

\begin{figure}
\includegraphics[width=8.5cm]{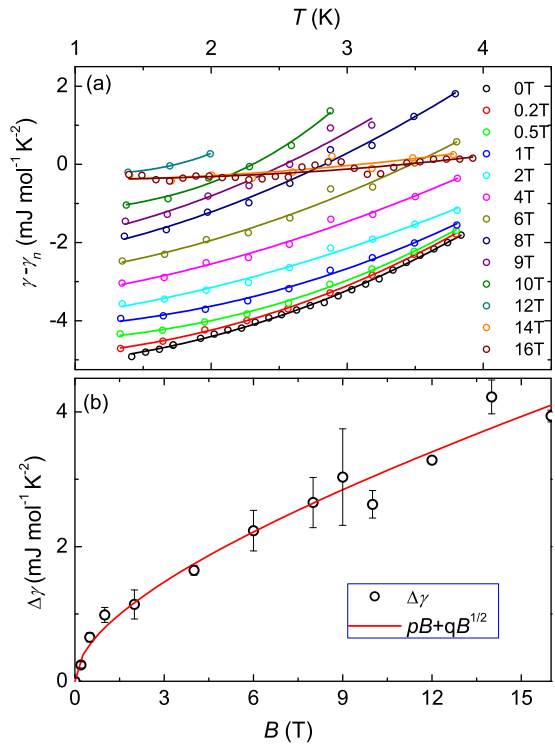}
\caption{\label{fig5} Magnetic field induced change of specific heat coefficient. (a) The raw data (symbols) and the fitting curves for each magnetic field. The fitting range varies depending on the region of the specific heat anomaly which should not be involved in deriving the low temperature intercept. (b) The magnetic field induced specific heat coefficient $\Delta\gamma$=[C(H)-C(0)/T]$|_{T\rightarrow0}$. The solid line shows a linear combination of two terms, namely $\Delta\gamma = p B + q B^{1/2}$.}
\end{figure}

\begin{table*}[htb]
\caption{\label{tab:table2}
The parameters derived from fitting to the angle dependence of the superconducting gaps}
\begin{ruledtabular}
\begin{tabular}{cccccccc}

 &$x_1(meV)$ & $x_2(meV)$ &$x_3(meV)$ & $x_4(meV)$ &  $x_5(meV)$ &$x_6(meV)$\\
\hline
$\alpha$ & 1.57 & 1.01  & -0.34 & 0.03 &  - &-\\
$\varepsilon$ & -1.27& 0.38   & 0.26 & 0.19 & 0.12 &0.03\\

\end{tabular}
\end{ruledtabular}
\end{table*}

Recently, Sprau et al. have done the STS measurements and used Bogoliubov quasiparticle interference imaging (BQPI) to measure the superconducting gap of FeSe \cite{Davis2016}. They proposed that the two superconducting gaps $\Delta_{\alpha}$ and $\Delta_{\varepsilon}$, which are located at hole pocket and electron pocket respectively, are both extremely anisotropic but nodeless. This highly anisotropic gaps are induced by the orbital selective pairing of the $d_{yz}$ orbital. In order to check the validity of their model, we use the gap structure determined by them to fit our specific heat data. Since the gaps are not the simple form of a sinusoidal function, therefore we need to fit their data first to an angle dependent function. The angle dependence of $\Delta_{\alpha}$ and $\Delta_{\varepsilon}$ from Ref.\cite{Davis2016} are plotted by the open symbols in Fig.~\ref{fig4}(a). The definition of the angle $\theta$ is illustrated in the inset of Fig.~\ref{fig4}(a). We try to fit each gap function with a combination of several harmonic of cosine functions, namely $\Delta=x_1+x_2cos(2\theta)+x_3cos(4\theta)+x_4cos(6\theta)+x_5cos(8\theta)+x_6cos(10\theta)$. The fitting results to the gaps are shown as red line and green line in Fig.~\ref{fig4}(a) and the fitting parameters are listed in Table~\ref{tab:table2}. With the expression of the gap structures, we obtain a remarkably good fit to the specific heat data in wide temperature region. In order to check whether the entropy is conserved about the low temperature anomaly, we also try to describe the second jump at 1.08 K by using an isotropic $s$-wave gap. This treatment is valid for the second anomaly either it is an AF or a superconducting transition. The blue line in Fig.~\ref{fig4}(b) shows the fitting with the gap structure in Fig.~\ref{fig4}(a) plus a small $s$-wave gap with $\Delta_s$ = 0.25 meV with $T_c$ = 1.08 K, and the fitting result is perfect. The fitting yields also the contribution of 32\% weight from the hole pocket and 68\% from the electron pocket. It seems that the gaps determined from the STS measurements can get good support from our specific heat data. The weights determined for different pockets are very important to further investigate the pairing mechanism of FeSe system.

\begin{table*}[htb]
\caption{\label{tab:table3}
The fitting parameters for superconducting electronic specific heat at different magnetic fields}
\begin{ruledtabular}
\begin{tabular}{cccccccc}

 magnetic field (T)&$a$ & $b$ &$n$ \\
\hline
 0 & -5.24 & 0.19 &2.16 \\
0.2 & -5.00 & 0.16 & 2.26\\
 0.5 & -4.59 & 0.11 & 2.48\\
 1  & -4.26 &0.13  &2.29 \\
  2  & -4.10 & 0.27 &1.77 \\
4  & -3.59 & 0.31 & 1.75\\
 6  & -3.01 & 0.29 &1.89 \\
 8  & -2.59 & 0.39 & 1.83\\
9   & -3.16 & 1.08 & 1.17\\
 10   & -1.27 & 0.09 &3.22 \\
12  & -0.3 &0.02  & 4.80\\
 14  & -0.45 & 0.03 & 2.35\\
 16   & -0.40 & 0.01 & 2.78\\

\end{tabular}
\end{ruledtabular}
\end{table*}

\subsection{Magnetic field dependence of specific heat in low temperature region}
In order to check whether the gaps have very large anisotropy, we have measured the specific heat of the FeSe single crystal at different magnetic fields. The superconducting electronic specific heat at different fields shown in Fig.~\ref{fig5} (a) are obtained by deducting the normal state contribution which has been derived above. Here we use the same data for the normal state for all fields supposing only phonon and fermionic contributions in the normal state and both are not dependent on magnetic field. By fitting the data with $\gamma = a + bT^n$ in the temperature window of about 1.36 K to about 3.8 K, we can get the superconducting electronic specific heat at 0 K at different magnetic fields. Here we must mention that, for the data at about 9 to 12 tesla, the superconductivity related anomaly or enhancement of specific heat has already come to this temperature region, therefore the fitting is done in much narrow temperature region, otherwise the fitting is invalid and would yield unreasonable results about the zero temperature intercept. Above about 12 T, the sample has already come to the normal state above 1.4 K. One can judge this by looking at the data at 14 and 16 tesla, which shows as a plain flat line. Through these treatments and fitting, we can have the magnetic field induced specific heat coefficient defined by $\Delta\gamma = [C(H) - C(0)]/T|_{T\rightarrow 0K}$. The fitting parameters are listed in Table~\ref{tab:table3}. The data are shown in Fig.~\ref{fig5} (b). We then fit the data $\Delta \gamma (H) =[C(H)-C(0)]/T$ with the formula  $\Delta\gamma = p B + q B^{1/2}$ with $p:q$ = 1 : 9 (B in unit of tesla), which seems working quite well for the data. As we know, the value of $\Delta\gamma$ depends on the magnetic field induced DOS at the Fermi level. When the superconducting gap is isotropic s-wave, the field dependence of $\Delta\gamma$ will follow a linear behavior, because only the vortex core part contributes the quasiparticles in the low temperature limit and the density of vortex linearly goes up with the magnetic field. If the gap structure is nodal like, for example d-wave, a square root relation will be observed due to the Doppler shift effect of the DOS outside the vortex core region \cite{G.E.Volovik1993JETP Lett}, which leads to a square-root temperature dependence of magnetic field. The relatively good fit with the formula $\Delta\gamma = p B + q B^{1/2}$  shown by the red solid line in Fig.~\ref{fig5} (b) does not manifest the $d$-wave gap, but suggests that the gap should have some deep minimum which leads to clear Doppler shift term. Thus the gaps should have anisotropic feature, and at least one gap should have the minimum which is close to zero.

\subsection{Concerning the specific heat jump at around 1.08 K}
Now let's address the possible origin of the specific heat jump at around 1.08 K. In the process of growing the crystal, we have used AlCl$_3$ as the flux because of its low melting temperature, there is a possibility that this specific anomaly is coming from the impurity of Al. If using the typical value of $\gamma_n$ and $T_c$ = 1.17 K for Al, we get a composition fraction of about 2$\%$ of Al in the sample. Although we could not exclude the possibility of Al as the contribution to this low temperature anomaly, we can however give several arguments to question that this may not be the case. The arguments are as follows: (1) This anomaly was also seen in similar shape in samples without AlCl$_3$ as flux \cite{Cava}. In that work, a similar anomaly of specific heat appears at around 1.25 K and it strongly depends on the concentration of interstitial Fe. (2) In another earlier study \cite{Lin.Jiao2016arXiv}, the authors observed a similar specific heat anomaly at around 2-3 K, this is much higher than the critical temperature of Al ($T_c$ = 1.17 K). (3) Most importantly, in our low temperature specific heat measurements, we find that this anomaly appears to a magnetic field at least up to 2000 Oe, but the critical field $H_c$ of Al is about 99 Oe. (4) Finally, from a chemical point of view, it is difficult to understand why AlCl$_3$ can easily separate into pure Al in the growing process. Therefore we would like to attribute this anomaly as the possible appearance of the long sought antiferromagnetic order. As far as we know, the AF order has never been observed in FeSe at ambient pressure, but it may exist with a very low Neel temperature. Therefore this AF order is very fragile and its appearance will depend on the subtle change of the properties of the sample, for example the concentration of interstitial Fe etc. \cite{Cava}. In this sense, the absence of the AF order in some samples does not imply that it will not happen in other samples, although the superconducting transition temperatures are quite close to each other. As far as we know, in previous neutron diffraction measurements \cite{ZhaoJunNatMat2015}, nobody has measured down to the temperature of about 1.08 K. And the usual magnetization measurement to detect the AF order does not work since it is hidden deeply in the superconducting state, and the magnetic screening due to Meissner effect or the vortex state can already cover all the signal from the AF order. Thus we would believe that this low temperature anomaly is most likely to be the AF transition. If this is true, the phase diagram of FeSe needs to be corrected.

\section{SUMMARY}
Specific heat down to 0.414 K under magnetic fields up to 16 T has been measured in FeSe single crystals. Variable kinds of gap functions have been tried to fit the data. It is found that a single gap, regardless of the gap functions (isotropic s-wave, anisotropic s-wave, and d-wave gap) cannot be used to fit the data. A combination of two gaps with at least a highly anisotropic gap can fit the data yielding the gap minimum of about 0.15 meV or smaller. We further find that the gaps determined by the recent STS experiment can describe the data perfectly. A second specific anomaly shows up as a jump of $C/T$ at around 1.08 K, it cannot be understood as the second superconducting gap of FeSe. This anomaly may be induced by the impurity of Al arising from the flux of AlCl$_3$. But a more reasonable picture based on several arguments would suggest that this anomaly is the long sought antiferrmagnetic transition which appears depending on the subtle change of the sample property.

\begin{acknowledgments}
We thank Christoph Meingast, Peter Hirschfeld, Igor Mazin, Ilya Eremin and Greg Stewart for helpful discussions. This work was supported by the National Key Research and Development Program of China (2016YFA0300401,2016YFA0401700), and the National Natural Science Foundation of China (NSFC) with the projects: A0402/11534005, A0402/11374144.
\end{acknowledgments}


\nocite{*}
\begin{thebibliography}{40}

\bibitem{WuMKPNAS2008} F. C. Hsu, J. Y. Luo, K. W. Yeh, T. K. Chen, T. W. Huang, P. M. Wu, Y. C. Lee, Y. L. Huang, Y. Y. Chu, D. C. Yan, and M. K. Wu, {\it Superconductivity in the PbO-type structure $\alpha$-FeSe}, Proceedings of the National Academy of Sciences of the United States of America {\bf105}, 14262 (2008).

\bibitem{Imai37K} S. Medvedev, T. M. McQueen, I. A. Troyan, T. Palasyuk, M. I. Eremets, R. J. Cava, S. Naghavi, F. Casper, V. Ksenofontov, G. Wortmann and C. Felser, {\it Electronic and magnetic phase diagram of bold italic $\beta$-Fe$_{1.01}$Se with superconductivity at 36.7 K under pressure}, Nature Materials {\bf8}, 630 (2009) .

\bibitem{Phase1Kothapalli} K. Kothapalli, A. E. B$\ddot{o}$hmer, T. Jayasekara, B. G. Ueland, P. Das, A. Sapkota, V.
Taufour, Y. Xiao, E. E. Alp, S. L. Bud'ko, P. C. Canfield, A. Kreyssig and A. I. Goldman, {\it Strong cooperative coupling of pressure-induced magnetic order and nematicity in FeSe}, Nature Commun. {\bf7}, 12728 (2016).

\bibitem{Phase2} J. P. Sun, K. Matsuura, G. Z. Ye, Y. Mizukami, M. Shimozawa, K. Matsubayashi, M. Yamashita, T. Watashige, S. Kasahara, Y. Matsuda, J. Q. Yan, B. C. Sales, Y. Uwatoko, J. G. Cheng, and T. Shibauchi, {\it Dome-Shaped Magnetic Order Competing with High-Temperature Superconductivity at High Pressures in FeSe}, Nat. Commun. {\bf7}, 12146 (2016).


\bibitem{Phase3} T. Terashima, N. Kikugawa, S. Kasahara, T. Watashige, Y. Matsuda, T. Shibauchi, and S. Uji, {\it Magnetotransport study of the pressure-induced antiferromagnetic phase in FeSe}, Phys. Rev. B {\bf93}, 180503(R) (2016)

\bibitem{BoehmerPRB2013} A. E. B$\ddot{o}$hmer, F. Hardy, F. Eilers, D. Ernst, P. Adelmann, P. Schweiss, T. Wolf, and C. Meingast, {\it Lack of Coupling between Superconductivity and Orthorhombic Distortion in Stoichiometric Single-Crystalline FeSe}, Phys. Rev. B {\bf87}, 180505(R) (2013).

\bibitem{TanatarPRL2016} M.A. Tanatar, A. E. B$\ddot{o}$hmer, E.I. Timmons, M. Sch$\ddot{u}$tt, G. Drachuck, V. Taufour, K. Kothapalli, A. Kreyssig, S.L. Bud¡¯ko, P.C. Canfield, R.M. Fernandes, and R. Prozorov, {\it Origin of the Resistivity Anisotropy in the Nematic Phase of FeSe}, Phys. Rev. Lett. {\bf117}, 127001 (2016).

\bibitem{ZhaoJunNatMat2015}Q. Wang, Y. Shen, B. Y. Pan, Y. Hao, M. W. Ma, F. Zhou, P. Steffens, K. Schmalzl, T. R. Forrest, M. Abdel-Hafiez, X. J. Chen, D. A. Chareev, A. N. Vasiliev, P. Bourges, Y. Sidis, H. Cao and J. Zhao, {\it Strong interplay between stripe spin fluctuations, nematicity and superconductivity in FeSe}, Nature Materials {\bf15}, 159 (2015).


\bibitem{BoehmerPRL2015} A. E. B$\ddot{o}$hmer, T. Arai, F. Hardy, T. Hattori, T. Iye, T. Wolf, H.V. L$\ddot{o}$hneysen, K. Ishida, and C. Meingast, {\it Origin of the Tetragonal-to-Orthorhombic Phase Transition in FeSe: A Combined Thermodynamic and NMR Study of Nematicity}, Phys. Rev. Lett. {\bf114}, 027001 (2015).



\bibitem{ShenZX} J. J. Lee, F. T. Schmitt, R. G. Moore, S. Johnston, Y.-T. Cui, W. Li, M. Yi, Z. K. Liu, M. Hashimoto, Y. Zhang, D. H. Lu, T. P. Devereaux, D.-H. Lee and Z.-X. Shen, {\it Interfacial mode coupling as the origin of the enhancement of $T_c$ in FeSe films on SrTiO$_3$}, Nature {\bf515}, 245 (2014).





\bibitem{DingH} P. Zhang, T. Qian, P. Richard, X. P. Wang, H. Miao, B. Q. Lv, B. B. Fu, T. Wolf, C. Meingast, X. X. Wu, Z. Q. Wang, J. P. Hu, and H. Ding, {\it Observation of Two Distinct $d_{xz}/d_{yz}$ Band Splittings in FeSe}, Phys. Rev. B {\bf91}, 214503 (2015).




\bibitem{Hanaguri} T. Watashige, Y. Tsutsumi, T. Hanaguri, Y. Kohsaka, S. Kasahara, A. Furusaki, M. Sigrist, C. Meingast, T. Wolf, H. v. L\"{o}hneysen, T. Shibauchi, and Y. Matsuda, {\it Evidence for Time-Reversal Symmetry Breaking of the Superconducting State near Twin-Boundary Interfaces in FeSe Revealed by Scanning Tunneling Spectroscopy}, Phys. Rev. X {\bf5}, 031022 (2015).



\bibitem{MatrusdaPNAS} S Kasahara\emph{et al.} S. Kasaharaa, T. Watashigea, T. Hanagurib, Y. Kohsakab, T. Yamashitaa, Y. Shimoyamaa, Y. Mizukamia, R. Endoa, H. Ikedaa, K. Aoyamaa, T. Terashimae, S. Ujie, T. Wolff, H. von L\"{o}hneysenf, T. Shibauchia, and Y. Matsuda, {\it Field-Induced Superconducting Phase of FeSe in the BCS-BEC Cross-Over}, Proc. Natl. Acad. Sci. USA, {\bf111}, 16309 (2014).


\bibitem{XueQK2010Science} C. L. Song, Y. L. Wang, P. Cheng, Y. P. Jiang, W. Li, T. Zhang, Z. Li, K. He, L. L. Wang, J. F. Jia, H. H. Hung, C. J. Wu, X. C. Ma, X. Chen, and Q. K. Xue, {\it Direct Observation of Nodes and Twofold Symmetry in FeSe Superconductor}, Science {\bf332}, 1410 (2011).



\bibitem{TaillefferPRL2016} P. Bourgeois-Hope, S. Chi, D. A. Bonn, R. Liang, W. N. Hardy, T. Wolf, C. Meingast, N. Doiron-Leyraud, and L. Taillefer, {\it Thermal Conductivity of the Iron-Based Superconductor FeSe: Nodeless Gap with a Strong Two-Band Character}, Phys. Rev. Lett. {\bf117}, 097003 (2016).


\bibitem{HardyBaK122}F. Hardy, T.Wolf, R. A. Fisher, R. Eder, P. Schweiss, P. Adelmann,
H. V. L$\ddot{o}$hneysen, and C. Meingast, {\it Calorimetric evidence of multiband superconductivity in Ba(Fe$_{0.925}$Co$_{0.075})_2$As$_2$ single crystals}, Phys. Rev. B {\bf81}, Phys. Rev. B {\bf81}, 060501(R)(2010).

\bibitem{MuGangBaK122}G. Mu, H. Luo, Z. Wang, L. Shan, C. Ren, and H.-H. Wen, {\it Low temperature specific heat of the hole-doped Ba$_{0.6}$K$_{0.4}$Fe$_2$As$_2$ single crystals}, Phys. Rev. B{\bf97}, 174501 (2009).

\bibitem{HardyK122}F. Hardy, A. E. B$\ddot{o}$hmer, D. Aoki, P. Burger, T. Wolf, P. Schweiss, R. Heid, P. Adelmann, Y. X. Yao, G. Kotliar, J. Schmalian, and C. Meingast,{\it Evidence of Strong Correlations and Coherence-Incoherence Crossover in the Iron Pnictide Superconductor KFe$_2$As$_2$}, Phys. Rev. Lett. {\bf111}, 027002 (2013).


\bibitem{J.Y.LinPRB2011} J. Y. Lin, Y. S. Hsieh, D. A. Chareev, A. N. Vasiliev, Y. Parsons, and H. D. Yang, {\it Coexistence of isotropic and extended s-wave order parameters in FeSe as revealed by low-temperature specific heat}, Phys. Rev. B {\bf84}, 220507(R) (2011).


\bibitem{Lin.Jiao2016arXiv} L. Jiao, C. L. Huang, S. R\"{o}{\ss}ler, C. Koz, U. K. R\"{o}{\ss}ler, U. Schwarz, S. Wirth, {\it Direct Evidence for Multi-Gap Nodeless Superconductivity in FeSe}, preprint at http://arxiv.org/abs/1605.01908.

\bibitem{Davis2016} P. O. Sprau, A. Kostin, A. Kreisel, A. E. B\"{o}hmer, V. Taufour, P. C. Canfield, S. Mukherjee, P. J. Hirschfeld, B. M. Andersen, J.C. S\'eamus Davis, {\it Discovery of Orbital-Selective Cooper Pairing in FeSe},  preprint at http://arxiv.org/abs/1611.02134.

\bibitem{G.E.Volovik1993JETP Lett} G. E. Volovik, {\it Superconductivity with lines of GAP nodes: density of states in the vortex} Jetp Letters {\bf 58}, 469(1993).

\bibitem{Cava} T. M. McQueen, Q. Huang, V. Ksenofontov, C. Felser, Q. Xu, H. Zandbergen, Y. S. Hor, J. Allred, A. J. Williams, D. Qu, J. Checkelsky, N. P. Ong, and R. J. Cava, {\it Extreme sensitivity of superconductivity to stoichiometry in Fe$_{1+\delta}$Se}, Phys. Rev. B {\bf79}, 014522 (2009).


\end{thebibliography}
\end{document}